# 1/$f$ Noise from the Laws of Thermodynamics for Finite-Size Fluctuations

Ralph V. Chamberlin and Derek M. Nasir,

Department of Physics, Arizona State University, Tempe, AZ 85287-1504

**Abstract**

Computer simulations of the Ising model exhibit white noise if thermal fluctuations are governed by Boltzmann's factor alone; whereas we find that the same model exhibits 1/$f$ noise if Boltzmann's factor is extended to include local alignment entropy to all orders. We show that this nonlinear correction maintains maximum entropy during equilibrium fluctuations. Indeed, as with the usual resolution of Gibbs' paradox that avoids net entropy reduction during reversible processes, the correction yields the statistics of indistinguishable particles. The correction also ensures conservation of energy if an instantaneous contribution from local entropy is included. Thus, a common mechanism for 1/$f$ noise comes from assuming that finite-size fluctuations strictly obey the laws of thermodynamics, even in small parts of a large system. Empirical evidence for the model comes from its ability to match the measured temperature dependence of the spectral-density exponents in several metals, and to show non-Gaussian fluctuations characteristic of nanoscale systems.

Low-frequency noise degrades many technologies [1-7], but it also defines the slow response of most materials [8-23]. Three types of noise were first measured about 90 years ago during the development of electronic amplifiers: shot noise, white noise, and 1/*f* noise [24-26]. Shot noise comes from the statistics of single-particle events. White noise comes from Gaussian fluctuations using standard Boltzmann statistics; whereas a general mechanism for 1/*f* noise has not yet been established. Here we present a fundamental mechanism for 1/*f* noise from non-Gaussian fluctuations when a Taylor-series expansion of local alignment entropy in Boltzmann's factor is extended to include the exact contribution from every configuration. One consequence is that this nonlinear correction makes nearby particles statistically indistinguishable, thereby ensuring extensive entropy and avoiding Gibbs' paradox [27]. Another result is that the nonlinear correction restores conservation of energy when an instantaneous contribution from the local entropy is included [28,29]. Finally, the nonlinear correction maintains maximum entropy during thermal fluctuations, thereby strictly preserving the $2^{nd}$ law of thermodynamics. Thus, a common mechanism for 1/*f* noise may be thermal fluctuations, similar to white noise but with a non-Boltzmann distribution characteristic of indistinguishable particles that interact on interatomic length scales. The mechanism applies when finite-size fluctuations contribute significantly to the equilibrium energy of the system.

The central-limit theorem yields Gaussian fluctuations for the properties of large systems with a well-defined mean value, but here we study small systems with large fluctuations that show non-Gaussian behavior on relatively short times, before the mean value is well defined. Other fluctuation theorems have given new insight into the behavior of systems that are far from equilibrium [30-32]. These theorems generally rely on the Boltzmann distribution, characteristic of systems that are (or were) weakly coupled to an effectively infinite heat bath. Here we study equilibrium fluctuations in small systems that may also be far from equilibrium. More importantly we include a type of reversible coupling between the system and its bath that influences the fluctuations.

We are guided by the principles of small-system thermodynamics that were developed to describe the behavior of individual molecules and isolated nanosystems [33]. We adopt these ideas to treat independently fluctuating regions inside bulk samples [27-29], as is found for the primary response of most materials [34-37]. A key feature of this "nanothermodynamics" is the subdivision potential, $\mathscr{E}$, which facilitates conservation of energy for finite-size systems. $\mathscr{E}$ can be understood by comparison to the chemical potential, $\mu$. $\mu$ is the change in energy to take a single particle from a bath of particles into the system, whereas $\mathscr{E}$ is the change in energy to take a cluster of interacting particles from a bath of clusters into the system. In general a cluster of $N$ interacting particles does not have the same energy as $N$ isolated particles due to: surface terms, length-scale effects, and thermal fluctuations. Thus $\mathscr{E}$ contains all non-extensive contributions to energy, including fluctuations in configurational entropy that do not couple linearly to the interaction energy in Boltzmann's factor.

We use Monte Carlo simulations of the Ising model to study thermal fluctuations. Although the Ising model was originally developed to describe ferromagnets, it remains one of the most widely used models for investigating thermal properties of interacting particles. A quadratic correction to Boltzmann's factor has been found to improve agreement between the Ising model and measured susceptibility of ferromagnetic materials and critical fluids [28], as well as nanometer-sized dynamical correlations in the structure of LaMnO$_3$ [29]. Here we extend the nonlinear correction to all orders, and show that it yields $1/f$ noise similar to many systems.

A useful interpretation of entropy is that it comes from missing information. Specifically, Boltzmann's entropy can be written as $S = k_B \ln(\Omega)$, where $k_B$ is Boltzmann's constant and $\Omega$ is the number of microstates that yield the observed macrostate. Here we study the alignment entropy of regions containing $n$ binary degrees of freedom ("spins"), where each spin may be up or down. If the alignment of every spin in a region is fixed, so that no information is missing, then $\Omega = 1$ and $S = 0$. If instead the region is fully isolated so that all information is missing, with no constraints on the alignment, then $\Omega = 2^n$ and $S = n\, k_B \ln(2)$. Between these extremes lies the usual alignment entropy of the Ising model, $S_m = k_B \ln\{n!/[½(n+m)]![½(n-m)]!\}$, found from the binomial coefficient for the number of ways that $n$ spins can yield the net alignment $m$.

Thermal fluctuations in a local region of a large sample are governed by the probabilities $w \sim e^{(\Delta S_m + \delta S^*)/k_B}$ [38]. Here $\delta S^*$ is the relatively small *change* in entropy of the bath, while $\Delta S_m = S_m - S_0$ is the *offset* in entropy of the region from its maximum value $S_0 = k_B \ln\{n!/[(½n)!]^2\}$. Boltzmann's factor $w \sim e^{-\delta E/k_B T}$ comes from the fundamental equation of thermodynamics for the bath at temperature $T$, $\delta S^* = \delta E^*/T$, with conservation of energy between the region and bath, $\delta E^* = -\delta E$. Gaussian fluctuations come from the lowest-order (quadratic) offset in entropy of the region, $\Delta S_m \propto -m^2$. Superficially these fluctuations might seem to violate the 2nd law of thermodynamics, but there are at least three possible explanations: 1) total entropy may decrease temporarily if the system is small enough [39]; 2) entropy should be calculated using Gibbs' ensembles that are independent of time; or 3) the entropy of the bath could increase to balance $\Delta S_m < 0$ in the region. Explanation 1) may apply to isolated systems, but here the region couples to its environment so that $S_m$ is not the total entropy. Explanation 2) suggests that the expression for entropy depends on the situation; $S = k_B \ln(\Omega)$ increases as a system evolves towards equilibrium, while Gibbs' formula avoids violating the 2nd law during thermal fluctuations [40,41]. Here we assume explanation 3): changes in entropy of the region are compensated by changes in entropy of the bath, thereby maintaining maximum entropy and retaining Boltzmann's definition. Thus, we assume that the entropy of the bath can be changed in two ways, from changes in energy and alignment of the region, so that successful inversion of a spin involves two criteria using random numbers between 0 and 1. The first criterion yields the Metropolis algorithm

$$e^{-\delta E/k_B T} > [0,1),  \quad\quad\quad \text{Eq. (1)}$$

where the step is accepted if Boltzmann's factor is larger than a random number between 0 and 1. The second criterion is the nonlinear correction to Boltzmann's factor

$$e^{(S_m - S_0)/k_B} > [0,1), \quad\quad\quad \text{Eq. (2)}$$

which yields $1/f$ noise in our model.

A similar (but not identical) nonlinear correction has been found to improve agreement between the Ising model and measured critical scaling in high-purity crystals [28]. One difference is that here we calculate the nonlinear correction in Eq. (2) to all orders using the exact expression for $S_m - S_0$, not just the lowest-order (quadratic) term. The other difference is that Eq. (2) was bypassed in [28] when $\delta E = 0$. Indeed, one reason for bypassing Eq. (2) was to avoid low-frequency fluctuations, whereas here we focus on these fluctuations. Another reason is that, unlike high-purity crystals $1/f$ noise may require defects [42], which reduce the likelihood of $\delta E = 0$ between states. $1/f$ noise can also be enhanced by non-equilibrium effects [43]. Here we present equilibrium fluctuations of the standard Ising model on simple-cubic lattices using either Eq. (1) or both Eq. (1) and Eq. (2) in every region, only briefly describing simulations where Eq. (2) is bypassed when $\delta E = 0$ in a subset of regions.

Our assumption that Eq. (2) comes from maintaining maximum entropy can be justified in other ways. One mechanism involves an additional change in energy of the bath from the offset in alignment entropy, as in adiabatic magnetization or demagnetization [44,45]. For $n$ non-interacting spins with magnetic moment $\mu_B$ in an external field $B$, the internal energy is [46] $E_m = TS_m - nk_BT \ln[2\cosh(\mu_B B/k_B T)]$. Here, the free energy (logarithmic term) comes from the thermal average over both states of each spin. Letting $B \rightarrow 0$ the free energy becomes constant, so that the offset in energy from its maximum is $\Delta E_m = T \Delta S_m$. This $\Delta E_m$ enhances the energy reduction when a region fluctuates into its low-entropy state, increasing the energy of the bath and furthering the fluctuation, consistent with Eq. (2). Another mechanism uses $S_m$ as a local bath of alignment entropy [28]. Specifically, high-entropy regions ($S_m \approx S_0$) have many states available facilitating fast spin flips; whereas low-entropy regions ($S_m \approx 0$) have few states available inhibiting spin dynamics, consistent with Eq. (2). Thus this mechanism is a type of entropic force, similar to Boltzmann's factor where the low entropy of a low-temperature bath inhibits transitions to higher energy. In any case, thermal fluctuations in small regions should obey small-system thermodynamics [33], which includes nonextensive thermal properties to all orders. Additional justification for Eq. (2) comes from the statistics of indistinguishable particles, as described below.

Figure 1 (a) depicts all possible alignments for a region containing two spins, $n = 2$. The left diagram shows that $S_{+2} = 0$ because there is only one way to have both spins up. Similarly, the right diagram has $S_{-2} = 0$. The middle diagram shows that there are two ways to have one spin up and the other spin down yielding $S_0 = k_B \ln(2)$, at least if the spins are distinguishable.

The dashed line in Fig. 1(b) indicates how the alignment entropy of the region might fluctuate as a function of time, between $S_0 = k_B \ln(2)$ for $m = 0$ and $S_m = 0$ for $m = \pm 2$. The dotted line shows how the entropy of the bath changes due to the nonlinear correction if total entropy is to remain maximized. Specifically, as the entropy of the region goes up-and-down, the entropy of the bath goes down-and-up, so that the total entropy of the region plus bath is constant (solid line).

Figure 1 (c) shows one consequence of using Eq. (2). When the entropy of the region is low the entropy of the bath is high, so that the aligned states tend to live longer. Specifically, the nonlinear correction favors aligned states when entropy is transferred to the bath, similar to how Boltzmann's factor favors low-energy states when energy is transferred to the bath. Here, the nonlinear correction causes each aligned state to live twice as long as Boltzmann's factor alone, so that each aligned state is as likely as both unaligned states. In general, as $T \to \infty$ where Eq. (1) can be ignored, Eq. (2) gives an average lifetime of each state $\tau_m \propto 1/e^{(S_m-S_0)/k_B}$, yielding a likelihood for each alignment that is independent of $m$ [28] $e^{S_m/k_B}\tau_m \sim e^{S_0/k_B} = n!/[(\tfrac{1}{2}n)!]^2$.

Figure 1 (d) shows an alternative interpretation of the nonlinear correction to Boltzmann's factor [28]. The central figure depicts how two configurations might be combined into a single $m = 0$ state, with each site containing a superposition of up and down spins, so that this unaligned state is as likely as each aligned state. For $n = 2$, these three alignments correspond to the triplet state of spin-½ particles, with the singlet state missing from this basic picture. Thus the nonlinear correction may be a simplistic way to simulate quantum-like statistics in an otherwise classical model. Indeed, a related nonlinear correction has been shown to restore extensive entropy and remove Gibbs' paradox from computer simulations of the Ising model [27,29], similar to how indistinguishable particles remove Gibbs' paradox in an ideal gas.

We simulate the Ising model on simple-cubic lattices of $N$ spins, with interaction energy $J$ between nearest-neighbor spins, and periodic boundary conditions on all external surfaces.

Large lattices are often subdivided into smaller regions, each containing $n \leq N$ lattice sites. Figure 2 (a) shows the total magnetization as a function of time, $M(t)$, from three sample sizes at two temperatures. Here $M(t)$ is multiplied by $\sqrt{N}$, scaling the amplitudes and showing that the sample-size dependence is consistent with thermal fluctuations. Note that the dynamics changes abruptly at $t = 0$. For $t < 0$ spin flips are governed by Boltzmann's factor alone (Eq. (1)), showing Gaussian fluctuations characteristic of white noise. For $t \geq 0$ the nonlinear correction (Eq. (2)) is added, yielding behavior characteristic of $1/f$ noise. Specifically, $M(t)$ shows large-amplitude wandering if the sample contains multiple regions (uppermost set of data); whereas $M(t)$ exhibits sharp jumps if the sample contains a single region (lower two sets of data). Similar on-off intermittency with varying time duration is known to yield $1/f$ noise in mathematical models [47,48]. In our model the jumps are due to entropic trapping from Eq. (2). Specifically, as the entropy of the region decreases (due to increased alignment) the entropy of the bath increases, enhancing the lifetime of these highly-aligned states. When the region eventually fluctuates back to high entropy, the time spent near $m=0$ is brief because the bath has low entropy.

Figure 2 (b) shows histograms from simulations of the $N = 12^3$ lattice (symbols), and from measurements on various systems (lines). Note that at high temperature our model yields trimodal behavior; from the maxima in entropy of the spins (central peak) and the bath (peaks near the endpoints). Similarly, fluctuations in a spin glass [15] and the ionic current through a nanopore [23] also yield significant probability at the center between the endpoints, unlike the purely bimodal behavior of a double-well potential [49].

Figure 3 shows power spectral densities as a function of frequency, $S(f)$, obtained from simulations of $M(t)$ similar to those in Fig. 2, but over much longer times for a wider frequency range. $M(t)$ is converted to the power spectral density using a discrete Fourier transform: $S(f) =$

$\left| \frac{1}{j} \sum_{t=0}^{j-1} M(t) \exp(-2\pi i f t / j) \right|^2$. The spectra are smoothed by linear regression using a sliding frequency range, where the spectral density at frequency $f_0$ comes from a linear least-squares fit to all data over the frequencies $-0.2 \leq 10 \log_{10}(f/f_0) \leq 0.2$. To obtain spectra over the entire frequency range without excessively large data files, we use a weighted average to combine independent simulations with different dwell times. Specifically, each simulation yields $2^{17} = 131,072$ data points, with dwell times of $10^0$ to $10^5$ sweeps between each data point. For convenience, all spectra are shifted so that $\log_{10}(f) = 0$ when $t = 10*2^{17}$ sweeps.

The solid sets of symbols in Fig. 3, which show nearly-constant spectral density (white noise), are from simulations using Eq. (1) alone. The open sets of symbols that show $1/f$-like behavior are from simulations using both Eq. (1) and Eq. (2). Over a wide range of frequencies these spectra are accurately characterized by $S(f) \propto 1/f^{\alpha(T)}$, with a temperature-dependent spectral-density exponent $\alpha(T)$.

Figure 4 shows $\alpha(T)$ as a function of $T/T_1$, where $T_1$ is the temperature at which $\alpha(T)$ extrapolates to 1. The solid symbols are from measurements [9] on various metallic films, given in the legend. The open symbols (connected by solid lines) come from simulations using both Eq. (1) and Eq. (2). These $\alpha(T)$ give the magnitude of the slope when plotted as in Fig. 3, determined by linear least-squares fits over one decade ($10 \log_{10}(f)=20-30$) with error bars from the standard deviation of three sets of simulations. Note that all simulations are at temperatures above the ferromagnetic transition, $k_B T/J > 20$ ($T/T_1 > 0.2$), and that this transition is much higher than for the Ising model using Eq. (1) alone ($k_B T/J \approx 4.5$) because highly-aligned states are favored by the increased entropy of the bath.

Figure 4 shows that $\alpha(T)$ decreases with increasing $T$ for measurements and simulations. A linear least-squares fit to the simulations at $T/T_1 \leq 1$ yields $\alpha(T) = 1.43 - 0.43\,(T/T_1)$ (dot-dashed line), showing good agreement with the temperature dependence of the measured $\alpha(T)$. Indeed, the slope of this least-squares fit (–0.43±0.02) is within experimental uncertainty of the average slopes from the four metallic films, –0.41±0.19. At higher temperatures, however, the simulations show $\alpha(T) \geq 1$ while the data have $\alpha(T) < 1$. An explanation may come from the fact that we find $\alpha(T) \leq 1$ for simulations with antiferromagnetic coupling between neighboring spins (not shown). Thus a more-detailed model that includes other interactions, such as dipolar fields or antiferromagnetic coupling between next-nearest neighbors, will be necessary to characterize the measured spectral-density exponents at high temperatures.

Although Fig. 3 shows $1/f$-like behavior over a wide range of frequencies, small regions exhibit saturation in their spectral density at low frequencies. Specifically, for both sets of data having $n=27$ the noise saturates below $10 \log_{10}(f) \approx -10$, remaining constant down to lowest frequencies. Thus these fluctuations have a well-defined mean value if averaged over long enough times. Assuming that the maximum number of steps for $1/f$ behavior comes from the maximum entropy of the region, $n!/[(\frac{1}{2}n)!]^2 = 2.04 \times 10^7$ steps for $n=27$. Indeed, for $n=27$ Fig. 3 shows that $1/f$ noise extends to about 7 orders of magnitude below the average attempt frequency. Similarly, smaller regions ($n = 8$, 12, and 18 spins, not shown) exhibit $1/f$ noise over smaller frequency ranges, consistent with $n!/[(\frac{1}{2}n)!]^2$. Furthermore, for $n = 64$ where $n!/[(\frac{1}{2}n)!]^2 = 1.83 \times 10^{18}$ steps, Fig. 3 shows no saturation over the full frequency range of our simulations, providing an explanation for the fact that saturation in $1/f$ noise at low frequencies is rarely observed in real systems. Thus, our model predicts a low-frequency limit to $1/f$ noise in every finite system, but because of the factorials in the entropy this limit can be at extremely low frequencies.

We have shown that a nonlinear correction to Boltzmann's factor yields 1/$f$ noise in simulations of a simple model, but most materials also show white noise at higher frequencies. One explanation is that many regions in real systems may have sufficient symmetry to yield nearly-degenerate states that cause Eq. (2) to be bypassed, as found for critical scaling in high-purity crystals [28]. Evidence that 1/$f$ noise involves defects comes from many measurements, including the dependence on electron irradiation [42]. The diamond-shaped symbols that lie below the other 1/$f$ spectra in Fig. 3 show simulations from a heterogeneous system, with one region using Eq. (1) and Eq. (2), while the other 26 regions have Eq. (2) bypassed when $\delta E = 0$. The combination of 1/$f$ noise at low frequencies and white noise at higher frequencies is similar to equilibrium measurements showing that both types of noise usually coexist [50]. Thus, both white noise and 1/$f$ noise may come from thermal fluctuations, with 1/$f$ noise requiring a nonlinear correction to Boltzmann's factor from the local entropy.

Detailed discussions with G. H. Wolf are gratefully acknowledged. We also thank S. Abe, N. Bernhoeft, B. F. Davis, P. D. Mauskopf, S. Moffet, N. Newman, P. F. Schmit, S. Seyler, and A. Shevchuk for helpful suggestions. Most of the simulations utilized the A2C2 computing facility. This research was supported by the ARO, W911NF-11-1-0419.

**Figure Captions**

**Fig. 1** (color online) Sketch of possible states in a two-spin region. (a) For distinguishable spins there is one way to have both spins up ($\Omega_{+2}=1$) or down ($\Omega_{-2}=1$), but two ways to have zero net alignment ($\Omega_0=2$). (b) During thermal fluctuations the Boltzmann entropy of the spins ($k_B\ln(\Omega)$) goes up and down (dashed line). To maintain maximum entropy we assume that the entropy of a thermal bath must go down and up (dotted line), so that the combined entropy of the system plus bath is constant (solid line). (c) When the bath has high entropy each low-entropy state in the region persists twice as long as expected from the Boltzmann factor alone. (d) Alternatively, zero alignment may come from a single state that contains a superposition of spins, consistent with de-localized particles that are indistinguishable in the region.

**Fig. 2** (color online) (a) Time sequence of magnetization per site from simulations on three lattice sizes at two temperatures, as given in the legend. Note that $M(t)$ is multiplied by $\sqrt{N}$ to scale the amplitudes, and the data from $N=12^3$ and $96^3$ are offset for clarity. At $t<0$ the spin-flip rate is governed by Boltzmann's factor alone, Eq. (1), yielding white noise. At $t\geq 0$ both Eq. (1) and Eq. (2) are used, yielding $1/f$-like noise. (b) Histograms from noise in simulations (symbols) and measurements (lines). Symbols are from the $N=12^3$ lattice, similar to (a) at $t \geq 0$ but over much longer time range. The top pair of lines comes from a spin glass at two temperatures [15]. The next line comes from ionic conduction through a nanopore [23]. The bottom pair of lines comes from a colloidal particle in two different double-well potentials [49].

**Fig. 3** (color online) Frequency dependence of spectral density (in dB) from simulations at $k_BT/J$=50 and 500, similar to those in Fig. 2. Note that $S(f)$ is multiplied by $N$ to scale different lattice sizes (given in the legend) and $\log_{10}(f)$ is multiplied by 10 to match the dB scale. Also note that the temperature dependence is relatively weak due to effective cancellation of the linear-$T$ dependence of thermal fluctuations and the nearly inverse-$T$ dependence of the magnetic susceptibility at high $T$. The spectra exhibiting white noise (bottom) come from using Eq. (1) alone. Spectra that exhibit 1/$f$–like behavior (diagonal) come from the same model using both Eq. (1) and Eq. (2). Over a broad range of frequencies these simulations can be characterized by $S(f) \propto 1/f^{\alpha(T)}$, with $\alpha(T) \approx 1.0$ for $k_BT/J$=500 (solid line) and $\alpha(T) \approx 1.15$ for $k_BT/J$=50 (dotted line). Diamond-shaped symbols, which show 1/$f$ noise at low frequencies and white noise at higher frequencies, come from a heterogeneous system described in the text.

**Fig. 4** (color online) Spectral-density exponent as a function of normalized temperature $T/T_1$, where $\alpha(T) \rightarrow 1$ as $T \rightarrow T_1$. Solid symbols are from measurements [9] on four metallic films. Open symbols connected by solid lines are from simulations using Eqs. (1) and (2). Specifically $\alpha(T)$ is the magnitude of the slope from simulations similar to those that lie along the diagonal in Fig. 3. The dot-dashed line shows a linear least-squares fit to $\alpha(T)$ from the simulations at $T/T_1 \leq 1$.

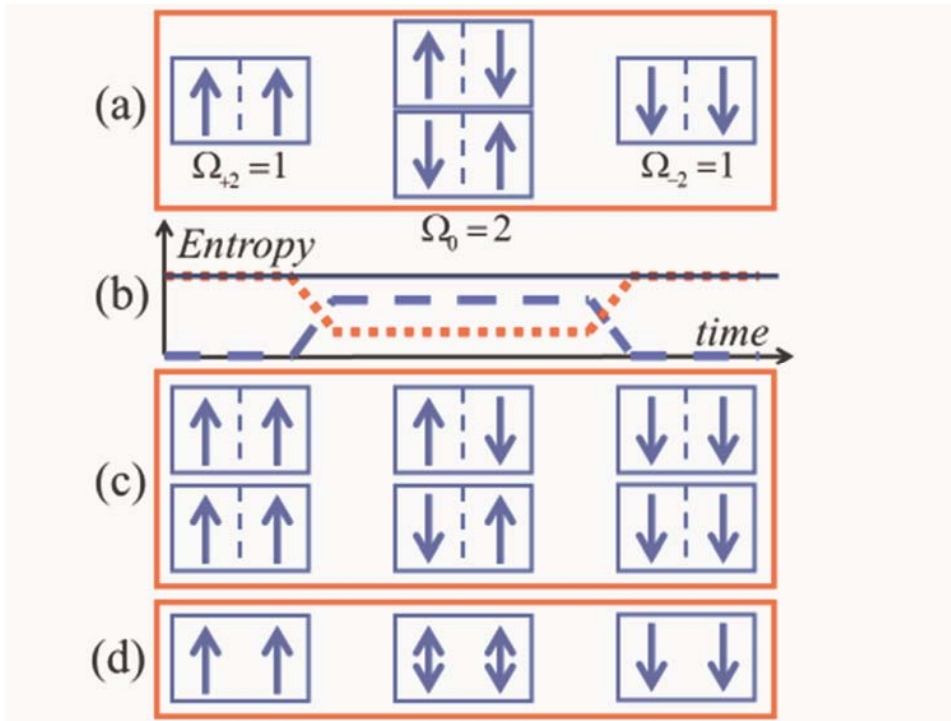

**Fig. 1**

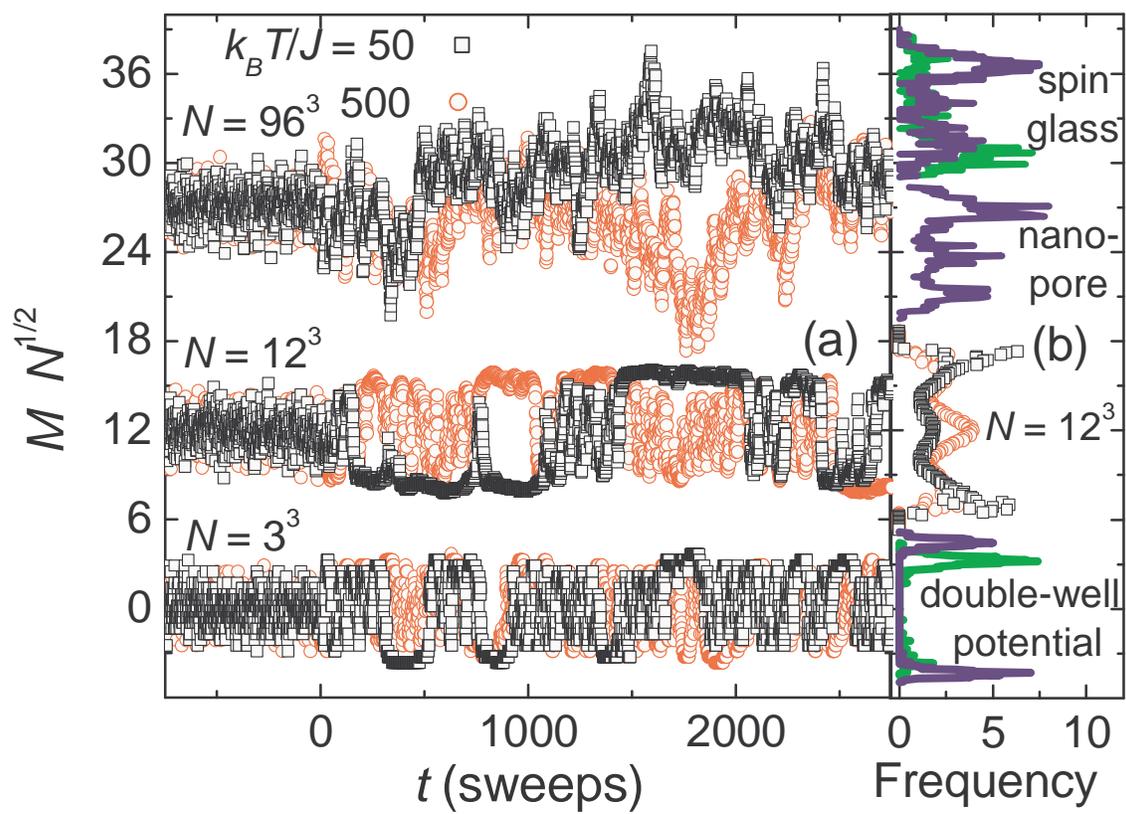

**Fig. 2**

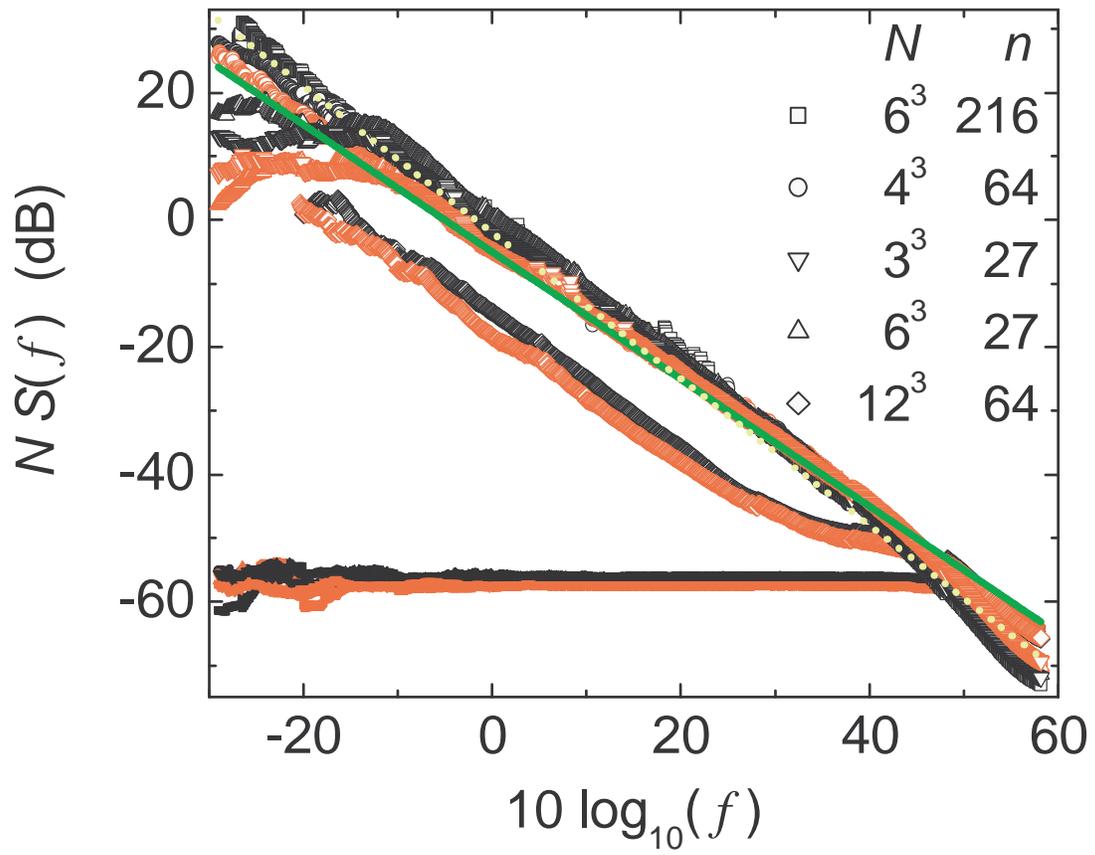

Fig. 3

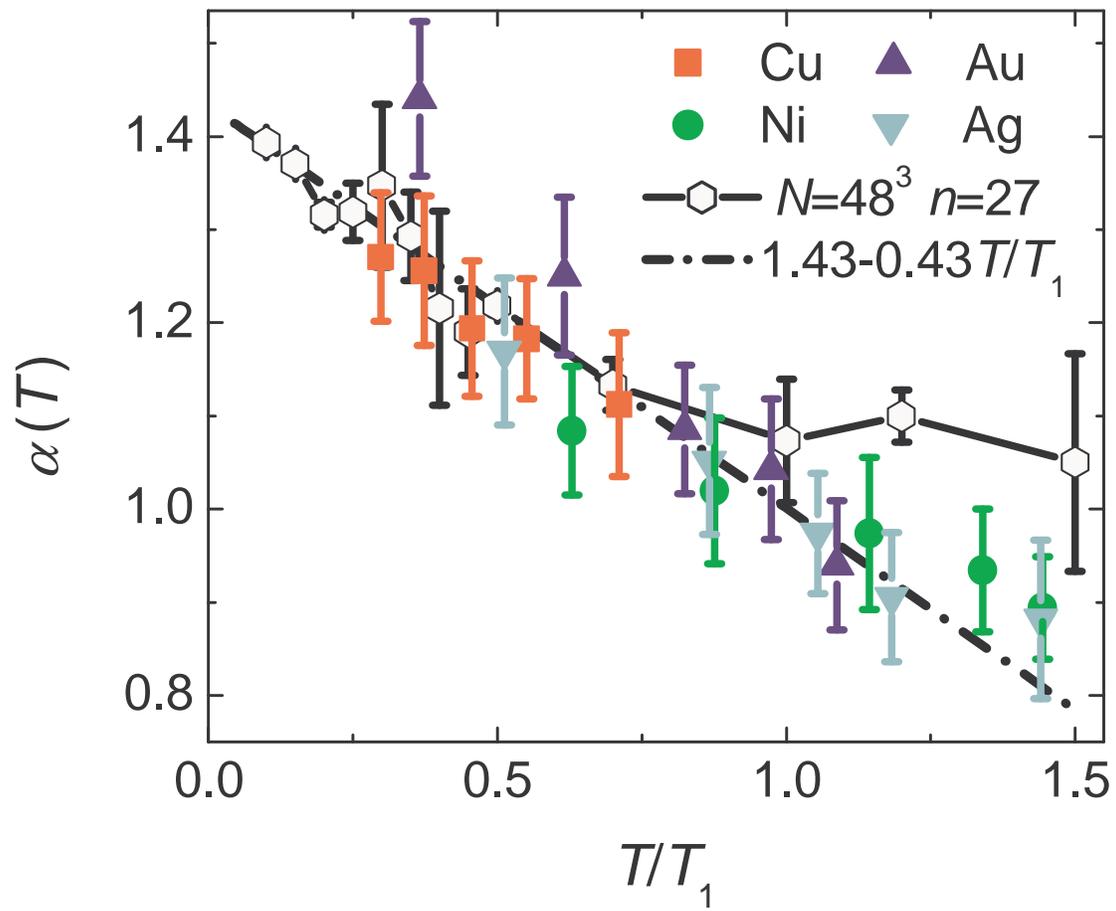

**Fig. 4**